\documentclass[12pt,preprint]{aastex}

\shorttitle{Observational Evidence for Coronal Twisted Flux Rope}
\shortauthors{N.-E. Raouafi 2008}

\begin{document}


\title{Observational Evidence for Coronal Twisted Flux Rope}


\author{N.-E. Raouafi\altaffilmark{*}}
\affil{National Solar Observatory, 950 N. Cherry Avenue, Tucson, AZ
85719, USA}

\altaffiltext{*}{Present address:  The Johns Hopkins University Applied Physics Laboratory, 11100 Johns Hopkins Rd, Laurel, MD
20723-6099, USA. E-mail: nour-eddine.raouafi@jhuapl.edu}



\begin{abstract}

Multi-instrument data sets of NOAA AR10938 on Jan. 16, 2007, (e.g., {\emph{Hinode}}, {\it{STEREO}},
{\it{GOES}}, {\it{MLSO}} and {\it{ISOON}} H$\alpha$) are utilized to study the fine structure and
evolution of a magnetic loop system exhibiting multiple crossing threads, whose arrangement and
individual shapes are very suggestive of individual field lines in a flux rope. The footpoints of
the magnetic threads are closely rooted into pores and plage areas. A C-class flare recorded by
{\it{GOES}} at approximately 2:35 UT near one of the footpoints of the multi-thread system (along
with a wisp of loop material shown by EUV data) led to the brightening of the magnetic structure
revealing its fine structure with several threads that indicate a high degree of linking
(suggesting a left-handed helical pattern as shown by the filament structure formed later-on). EUV
observations by {\emph{Hinode}}/EIS of hot spectral lines at 2:46~UT show a complex structure of
coronal loops. The same features were observed about 20 minutes later in X-ray images from
{\emph{Hinode}}/XRT and about 30 minutes further in EUV images of {\it{STEREO}}/SECCHI/EUVI with
much better resolution. H$\alpha$ and 304~{\AA} images revealed the presence of several filament
fibrils in the same area. They evolved a few hours later into a denser structure seemingly showing
helical structure, which persistently lasted for several days forming a segment of a larger scale
filament. The present observations provide an important indication for a flux robe as a precursor
of a solar filament.

\end{abstract}



\section{Introduction}

Most solar atmospheric features with varying degrees of complexity (i.e., active regions,
prominences, filaments, loops, etc.) are thought to be shaped by magnetic fields emerging from the
solar interior due to the buoyancy instability \citep{Parker84}. Complex topologies (twist, writhe,
linking and shear) of emerging flux tubes are key ingredients for numerous theoretical studies of
solar eruptions resulting from magnetic reconnection \citep[prominences, filaments, flares;
see][]{Linton98,Linton01}. In fact, highly twisted flux tubes store magnetic energy that
is necessary for the heating and acceleration of the plasma in the erupting structures. X-ray
observations have shown that the majority of the active regions leading to coronal mass ejections
(CMEs) have an S-shaped structure \citep{Canfield99}. \citet{Amari00} studied the crucial role
played by twisted flux tubes in the formation of topologically complex flux ropes and their
evolution into solar eruptive phenomena, such as CMEs.

Simulations of flux emergence from the convective zone into the upper atmosphere showed that
untwisted flux tubes suffer convective stresses leading to their fragmentation and impeding their
emergence to the upper atmosphere \citep{Schussler79,Longcope96}. It has been found that some
degree of twist in rising flux tubes is needed to avoid the conversion of the tube into vortex
pairs \citep{MoreEmon96}. It has also been reported that the flux reaching the upper atmosphere
depends on both the field strength at the bottom of the convective zone (simulation box) and the
degree of twist of the rising flux tube \citep{MartSyk08}.

Different mechanisms have been invoked to be at the origin of the twist of rising flux tubes: (1)
Helical turbulent motions \citep{Longcope98}; (2) Coriolis force \citep{FanGong00}; (3)
Differential rotation \citep{DeVore00}; (4) Helicity generation by the solar dynamo
\citep{Seehafer03}; (5) Turbulent diffusion of wrapped poloidal flux into the rising flux tube
\citep{Chatterjee06}.

Measurements of the magnetic field vector in sunspots have shown that the magnetic field has a
preferential helicity sign in both hemispheres \citep[negative in the north and positive in the
south; see][]{Seehafer90,Pevtsov95,Pevtsov01}. \citet{Lites95} studied the topology of the emerging
flux tube in a $\delta$-sunspot and found a rather simple structure at the start with increasing
topological complexity. \citet{Leka96} found that proper motions imply that flux bundles are
twisted before they emerge.

In the last few decades particular attention was paid to coronal filaments and prominences, which
can erupt into CMEs. At the chromospheric level, they are characterized by a channel of fibrils
with the filament spine, if present, laying above. An EUV or X-ray arcade of loops forms a
``transverse'' dome relative to the filament spine with a cavity separating both structures
\citep[see][for a review]{Martin98}. Numerous studies were dedicated to characterize the formation
and evolution of these structures.  Resolving the fine structure of these structures is important
to understand the mechanisms leading to their eruption and to constrain models. \citet{Li98}
studied fine structure of filaments through the interpretation of spectroscopic observations. They
found evidence for the presence of two dynamically different threads with different thermodynamic
properties. \citet{Pojoga98} compared prominence spectra to models taking into account radiative
transfer effect. They found different structures with different optic opacity along the line of
sight. Although resolving observationally the fine structure of filaments and prominences proved to
be difficult to achieve, \citet{Chae00} used EUV observations to study qualitatively the chirality
of filaments through the crossing topology of bright and dark threads.

The active region NOAA AR10938 is the area of interest for the present study. On Jan. 16, 2007, it
was located approximately at N02E30. Multi-instrument observations, mainly from the {\emph{Hinode}}
\cite[][]{Kosugi07} and the {\emph{Solar TErrestrial RElations Observatory}}
\cite[{\emph{STEREO}}:][]{Kaiser08}  missions, are utilized to study the formation and evolution of
a loop system that is highly suggestive of a flux rope.

\section{Observations}

The {\emph{Hinode}} Extreme UV Imaging Spectrometer \citep[EIS:][]{Culhane07} carried-out
three raster sequences of AR10938 on Jan. 16, 2007, at 1:54~UT, 2:20~UT and 2:46~UT with a
{1\arcsec} slit. The observations were acquired in a number of spectral lines whose formation
temperatures span a range from $\sim 0.08$ to $> 15$~MK.

The {\emph{Hinode}} Solar Optical Telescope \citep[SOT:][]{Tsuneta08} filtergram (hereafter SOT-FG)
and X-Ray Telescope \citep[XRT:][]{Golub07} were observing the active region jointly, within the
same time intervals and with comparable temporal cadence, on Jan. 15-16, 2007. High-resolution
LOS-magnetograms from SOT-FG are utilized to study the photospheric evolution of magnetic flux
(emergence and dynamics). The data is recorded on Jan. 15, 2007, 10:57 - 15:51 UT and 22:18 UT
until Jan. 16, 2007, 5:59 UT with a temporal cadence of approximately 1 minute. XRT observations
(Al-poly filter: $\log T\approx5.5 - 8.0$, with maximum around 6.9) along with
{\it{STEREO}}/SECCHI/EUVI \citep[][hereafter EUVI]{Howard08} data provide a proxy of the topology
of the different coronal loop systems.

EUVI-A was recording with a time cadence of 10 minutes in both 171~{\AA} and 195~{\AA}, while
EUVI-B was observing hourly. Since these observations were recorded soon after the {\it{STEREO}}
launch, the angular separation of the two satellites was very small and 3D reconstruction of the
observed structures is not possible. Thus, we limit ourselves to EUVI-A observations. EUVI-A
304~{\AA} images, which are taken with a lower time cadence, are also used to study cool
counterpart structures in relation with X-ray and EUV ones.

Additional data from other instruments ({\it{GOES}}, {\it{SOHO}}/EIT, {\emph{ISOON}}) are also
utilized to acquire complementary informations on the activity level of AR10938. They are, however,
not presented here.

\section{Results}

\begin{figure*}
\includegraphics[height=0.8\textheight]{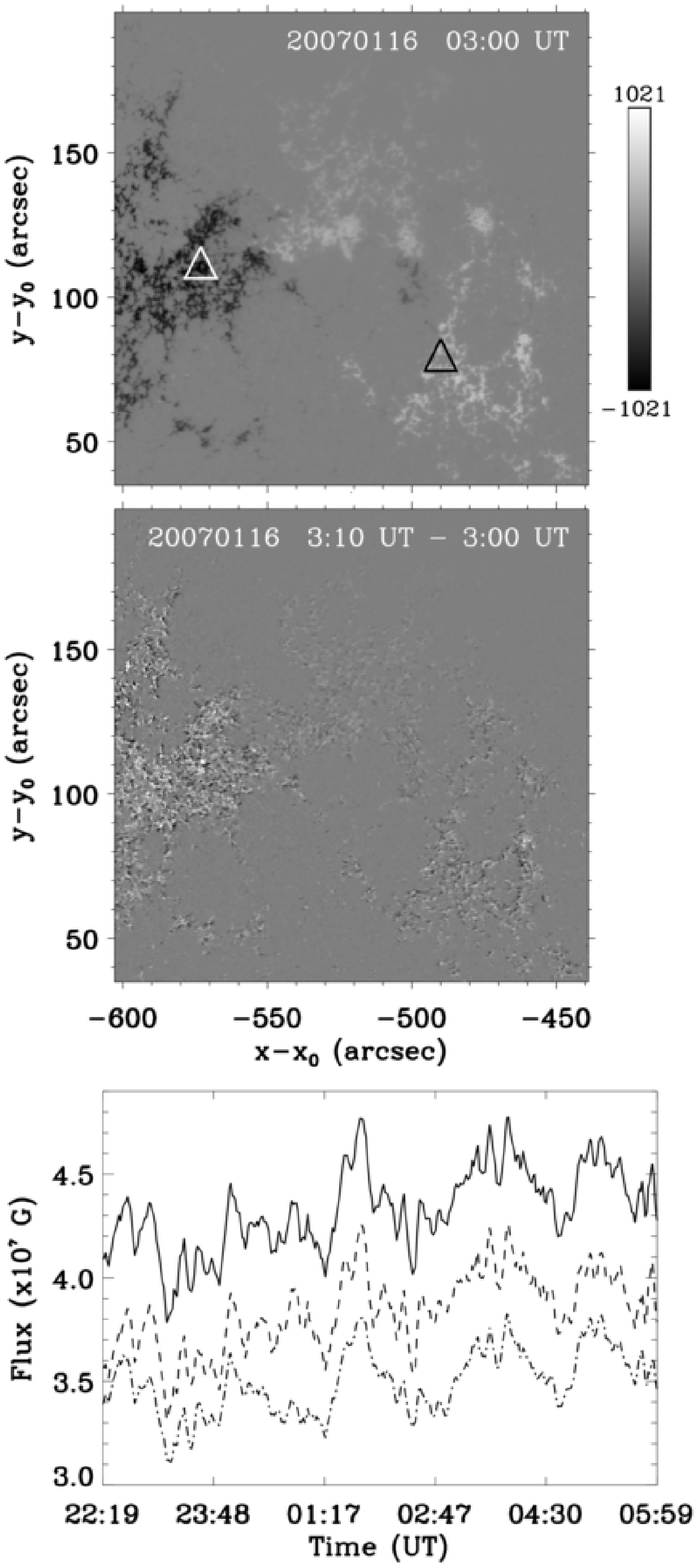}
\caption{Top: LOS-magnetogram from {\emph{Hinode}}/SOT-FG. The triangles indicate the location of the footpoints
of the twisted loops. Middle: difference map of the unsigned of magnetic
flux between 3:10 UT and 3:00 UT Jan. 16, 2007. Bottom: changes in the total flux (solid line),
negative polarity ($\times1.4$; dashes) and positive polarity ($\times2.2$; dot-dashes) as a function
of time. \label{hinode_sotfg}}
\end{figure*}

Fig.~\ref{hinode_sotfg} displays a LOS-photospheric magnetogram from SOT-FG of AR10938. The
triangles depict the footpoints' locations of the different threads forming the magnetic structure
suggesting the presence of the twisted coronal flux rope. The different threads are rooted in pores
and plage areas at both ends. The middle panel shows a difference map of the unsigned flux. It is
clear that significant changes occur in the regions near the loops' footpoints. For instance, the
large changes occur within the positive and negative polarity regions in the left-central area and
toward the bottom-right corner of the map, respectively. Flux changes elsewhere in the map are not
as important. This trend persisted for several hours. The bottom panel exhibits the temporal
evolution of the magnetic flux. Similar variations are obtained for the areas where the important
changes occurred. The flux fluctuations are indicative of changes in the magnetic field topology in
particular in the areas where the threaded magnetic structure is rooted. Starting from 23:00 UT, an
overall increase in the total flux (and also that of both polarities) is found until approximately
1:30 UT when a significant decrease coincided with increasing coronal activity.

{\emph{GOES}} recorded 12 X-ray bursts that occurred in AR10938 between 14:00 UT Jan. 15, 2007, and
16:00 UT Jan. 16, 2007. The most prominent one (a C-class flare) occurred on Jan. 16 at
approximately 2:35 UT. White light coronagraphs {\emph{STEREO}}/SECCHI/COR1 \& COR2
\cite[][]{Howard08} and {\emph{SOHO}}/LASCO \cite[][]{Brueckner95} did not detect any CME material
in relation with the C-class flare. On the other hand, EUVI images show a material wisp very likely
in conjunction with the flare.

EIS raster sequence performed at 2:46 UT reveals enhanced emissions in hot lines (Fe~{\sc{xxiv}}
255.1~{\AA}: $\log T=7.2$;  Ca~{\sc{xvii}} 192.82~{\AA}: $\log T=6.7$; Fe~{\sc{xvi}} 262.98~{\AA}:
$\log T=6.4$; and Fe~{\sc{xv}} 284.16~{\AA}: $\log T=6.3$). These emissions show a relatively
complex system of loops with NE-SW direction (footpoint locations are indicated in
Fig.~\ref{hinode_sotfg}).

\begin{figure*}
\includegraphics[width=0.95\textwidth]{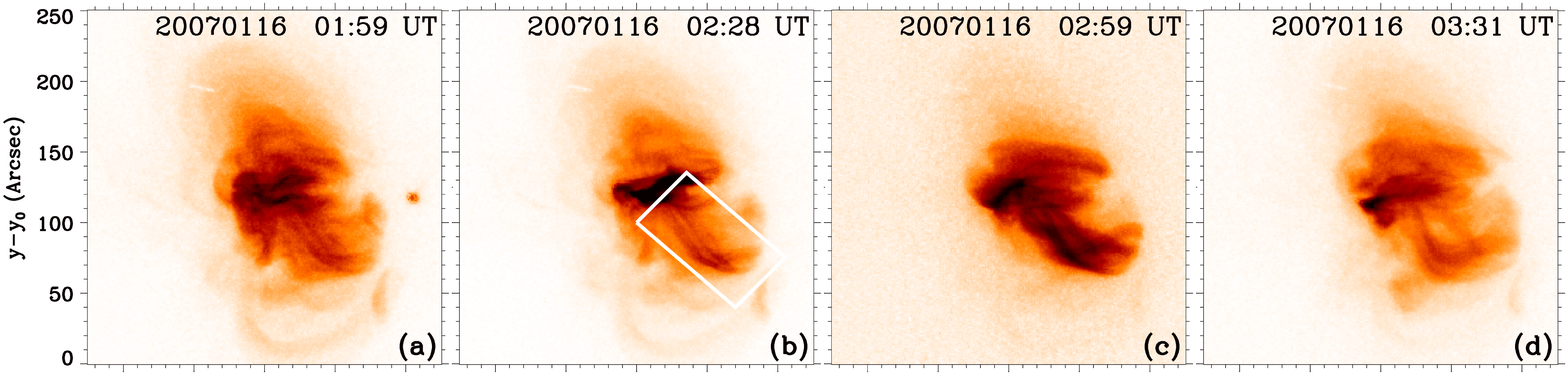}
\includegraphics[width=0.95\textwidth]{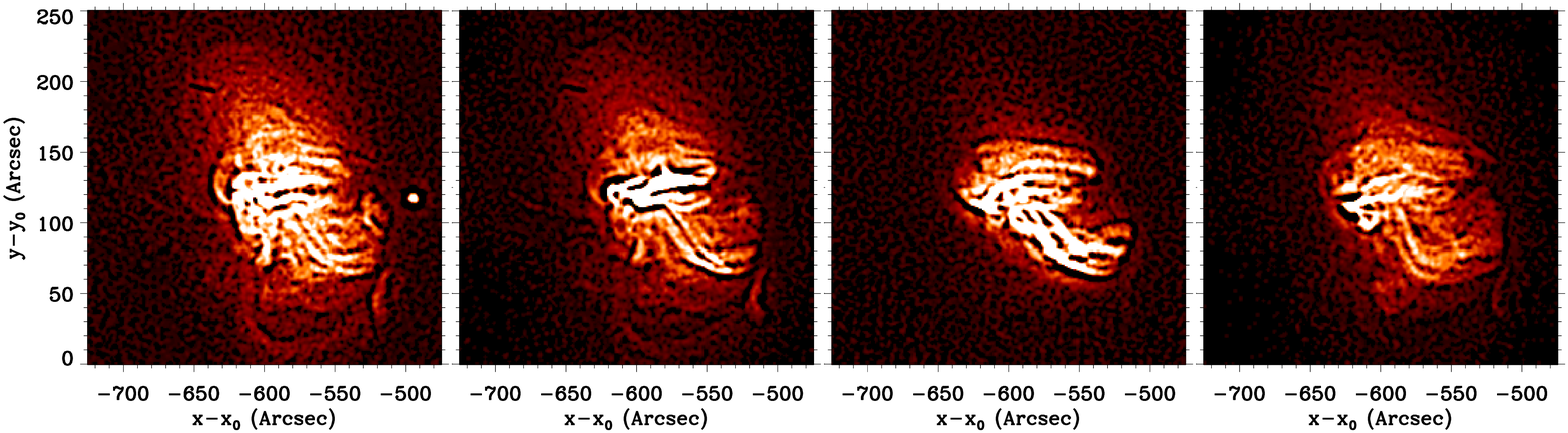}
\caption{Top: {\emph{Hinode}}/XRT snapshots of AR10938 showing the brightening of the
conjectured flux rope within the white box after the C-class flare at 2:35 UT ({\it{GOES}}).
Bottom: Same as above after contrast enhancement by wavelet filtering. \label{hinodexrt}}
\end{figure*}

The X-ray data recorded simultaneously with the SOT-FG photospheric magnetograms shows rapidly
evolving coronal loop structures. Top panels of Fig.~\ref{hinodexrt} display snapshots illustrating
the development of active region AR10938. The bottom panels show the same structures with enhanced
contrast after application of a wavelet filtering as described by \citet[][]{Stenborg03}. Different
loop systems are expanding rapidly in the corona with relatively simple topology, i.e. loops toward
the bottom of the different panels of Fig.~\ref{hinodexrt}. However, the bright loop system within
the white box in Fig.~\ref{hinodexrt}b is of particular interest. It had an apparent simpler
topology shown by X-ray data recorded earlier (10:57~UT - 15:51~UT Jan. 15, 2007). The occurrence
of the structure within the white box (Fig.~\ref{hinodexrt}b), that is extending from northeast
(-585\arcsec,120\arcsec) into southwest (-515\arcsec,80\arcsec), preceded the brightening of
several other loops (see Fig.~2c) forming a rather complex pattern. The features of the X-ray
system compare relatively well to the ones observed in hot line emissions observed by EIS.

\begin{figure*}
\includegraphics[height=0.45\textheight]{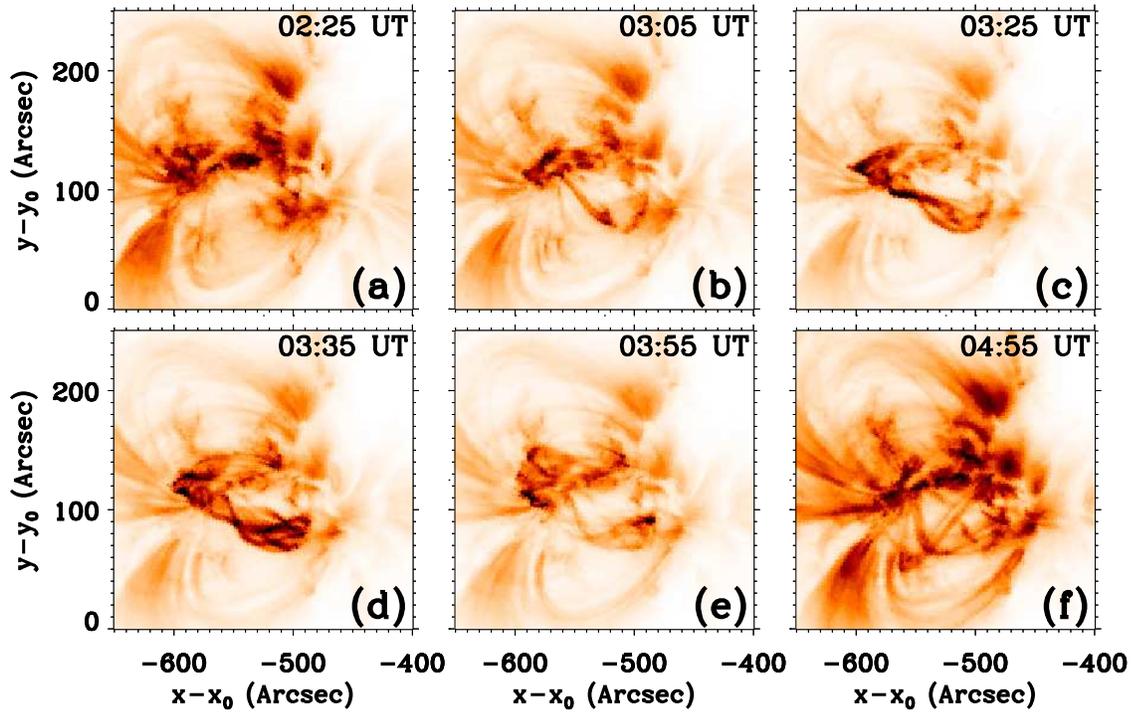}
\caption{171~{\AA} images from EUVI-A on Jan. 16, 2007, illustrating the evolution of the EUV
counterpart of the X-ray threads observed by {\emph{Hinode}}/XRT. The EUV structures look
similar to those observed in X-rays with, however, a time delay greater than 30 minutes in
appearance. \label{secchi_euvia}}
\end{figure*}

EUVI-A 171~{\AA} (Fig.~\ref{secchi_euvia}) and 195~{\AA} images recorded roughly between 3:00 UT
and 4:00 UT show the relatively cooler ($\sim1$~MK) counterpart of the loop system observed by EIS
and XRT (see Fig.~\ref{hinodexrt}). The system developed rapidly within a time interval of
about 30 minutes as shown by Figs.~\ref{secchi_euvia}b-\ref{secchi_euvia}d.

Fig.~\ref{secchi_euvia}d shows the fully developed complex topology of the threaded system. A
number of the new EUV loops do not correspond necessarily to the ones observed earlier as was the
case of the X-rays with respect to the EIS ones. The topology of the different threads indicates a
high degree of linking suggesting the presence of a twisted flux rope (see
Fig.~\ref{secchi_euvia}c-d). Fig.~\ref{secchi_euvia}c displays bright loops seemingly with
left-handed helical pattern as suggested by the presence of an inverse S-shaped filament \citep[see
Fig.~4e;][]{RustMartin94}. The system evolved further as a number of loops dimmed and disappeared
later. Other seemingly higher loop systems brightened later as shown by Fig.~\ref{secchi_euvia}f.
These show some similarities with X-ray loops that appeared earlier (see Fig.~\ref{hinodexrt}d).

The sequential appearances of the suggested flux rope in hot emission lines (e.g., Fe~{\sc{xxiv}}
225.1~{\AA}, $\log T=7.2$), then in X-ray images (a few MK), and finally in EUV 171~{\AA} and
195~{\AA} images ($\sim1$~MK) show the gradual cooling of the magnetic structure and display
details of its fine structure. A sheared arcade of loops is also seen seemingly crossing from above
the indicated flux rope in 171~{\AA} and 195~{\AA} images (Fig.~\ref{secchi_euvia}f). Similar
arcade was also seen prior to the flare and the appearance of the magnetic thread system.

\begin{figure*}
\includegraphics[height=0.45\textheight]{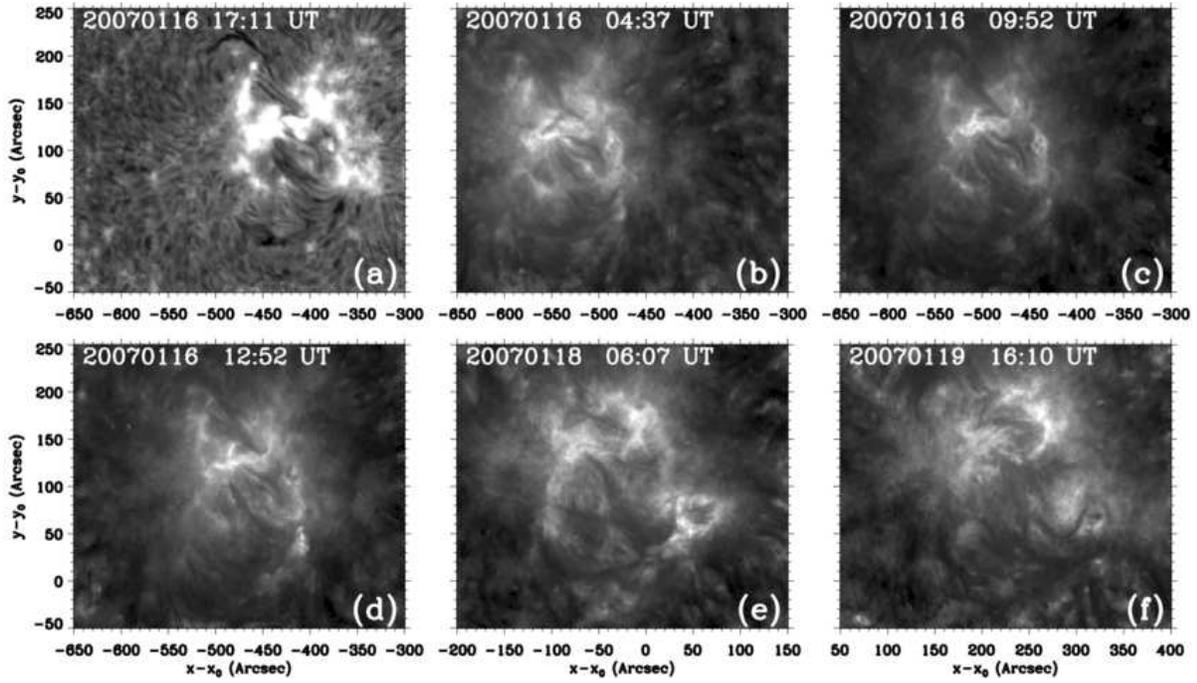}
\caption{(a) H$\alpha$ from Mauna-Loa Solar Observatory. The filament along the neutral line of
AR10938 is well developed north the active region and is of filamentary nature south of it. (b-f)
304~{\AA} images from {\it{STEREO}}/SECCHI/EUVI-A illustrating the evolution of the filament segment
corresponding  to the thread system seen in EUV and X-rays.\label{secchi_euvia_304}}
\end{figure*}

EUVI-A 304~{\AA} images reveal the presence of fibrils along the path of the magnetic thread
structure observed in EUV and X-ray data. These fibrils, also seen in H$\alpha$ from Mauna Loa
Solar Observatory and {\it{ISOON}}-Sac-Peak, were present several hours before the appearance of
the multiple crossing thread system. These are probably part of a larger structure forming an
inversed S-shape (see Fig.~\ref{secchi_euvia_304}a-b) extending along the neutral line across
AR10938. This is well developed north to the active region and is of filamentary nature within and
south of the same active region. 304~{\AA} data reveal a denser structure along the location of the
indicated flux rope that formed few hours later with an apparent helical pattern (see
Fig.~\ref{secchi_euvia_304}d). The presence of the filament was more prominent the next few days,
in particular on Jan. 18, 2007, (see Fig.~\ref{secchi_euvia_304}e-f). We believe this is a segment
of the larger filament structure.  H$\alpha$ images showed also the presence of prominence activity
coinciding with the presence of the active region at the west limb on Jan. 24-26, 2007.

\section{Conclusions and Discussion}

Multi-instrument observations of the active region NOAA AR10938 on Jan. 16, 2007, provide evidence
for a magnetic thread system with a complex, multiple crossing topology, which is highly suggestive
of a flux rope. A C-class flare in the active region AR10938 led to the brightening of the magnetic
structure showing its fine structure in an unprecedented manner. The fine structure of the magnetic
system is best seen in XRT images at approximately 3:00 UT and with better contrast in EUVI-A
171~{\AA} and 195~{\AA} data at about 3:30 UT. It is very unlikely that the observed complex
topology is the result of projection effects. XRT data suggests that the indicated flux rope was
present prior to the flare and its appearance is the result of heating processes related to the
flare eruption. This led to emissions in hot spectral lines observed by EIS, followed by a cooling
phase through X-rays and EUV, and ultimately the gradual disappearance of the system.

H$\alpha$ and 304~{\AA} data reveal the presence of dark fibrils aligned along the suggested flux
rope. These structures were present before the appearance of the magnetic thread system. The X-ray
and EUV threads suggest a high degree of linking as shown by the formation of a segment of an
inverse S-shaped filament in the same location later-on conveying a left-handed helical pattern for
the loop system. We believe that the dark fibrils are part of a filament channel and the indicated
flux rope is a segment of the filament. The latter runs along the neutral line across the active
region. It is well developed north of AR10938 and of filamentary nature elsewhere. This is
supported by the presence of a loop arcade presumably laying above the multi-thread system, which
is necessary for the formation of filaments. 304~{\AA} observations show a dark, denser structure
that formed about 8 hours after the brightening of the suggested flux rope. This evolved further
into a wider feature in the following days reflecting the formation of the filament which is also
supported by the presence of prominences when the active region was near the solar limb.

The present study should be useful for constraining models of filament formation. The role of the
magnetic field topology, in terms of twist and shear, is a matter of debate concerning the
processes of the filament formation. An important aspect is how and when (with respect to the
eruptive phase of filaments) the confinement of the magnetic fields occurs. Two types of models are
proposed to address these questions. The first class is flux rope based models which assume a high
degree of twist from the start of solar filaments' formation
\citep{RustKumar94,PriestForbes90,Low01}. The second class considers highly sheared arcades along
magnetic inversion lines to be the base of the filament, where the helical structure of magnetic
field occurs only during the eruptive phase for the latter type of models
\citep{Pneuman83,vanBallegMart89,vanBallegMart90,Antiochos94}. Although the present work is not
meant to discriminate between the two classes of models, it indicates the presence of a flux rope
prior to the filament formation. We believe that it favors the first class. The processes of the
plasma condensation in the loop system leading to the filament formation remain unclear.

It is likely that the flux tube emerged twisted from the convection zone. Photospheric shear
motions may also contribute to the twist of the flux tube. However, shear transfer during the
flare, which led to the brightening of the thread system, may be more plausible. The shear within
the non-eruptive flaring structure should remain within the system. The only plausible way for this
to happen is to increase the topological complexity of neighboring magnetic structures. A more
detailed study of the dynamics within the active region is needed and will be carried out in the
future.

\acknowledgments

NSO is operated by the AURA, Inc., under cooperative agreement with the NSF. {\emph{Hinode}} is a
Japanese mission developed and launched by ISAS/JAXA, with NAOJ as domestic partner and NASA and
STFC (UK) as international partners. It is operated by these agencies in co-operation with ESA and
NSC (Norway). The {\it{STEREO}}/SECCHI data used here are produced by an international consortium
of the NRL (USA), LMSAL (USA), NASA GSFC (USA), RAL (UK), Univ. Birmingham (UK), MPS (Germany), CSL
(Belgium), IOTA (France), and IAS (France). N.-E. R.'s work is supported by NASA grant NNH05AA12I.

\clearpage



\end{document}